\documentclass[11pt]{article}

\usepackage{acl}

\usepackage{times}
\usepackage{latexsym}

\usepackage[T1]{fontenc}

\usepackage[utf8]{inputenc}

\usepackage{amsmath}
\usepackage{booktabs}
\usepackage{tabularx} 
\usepackage{xcolor}
\usepackage{colortbl}
\usepackage{enumitem}
\usepackage{amssymb}
\usepackage{multirow}
\usepackage{subcaption}
\usepackage{soul}

\usepackage{microtype}

\usepackage{inconsolata}

\usepackage{graphicx}

\title{
Decision-aware User Simulation Agent \\ for Evaluating Conversational Recommender Systems
}

\author{Yuan-Chi Li \quad Li-Chi Chen \quad Sung-Yi Wu \quad Yu-Che Tsai \quad Shou-De Lin \\
  Department of Computer Science and Information Engineering \\
  National Taiwan University \\
  \texttt{\{b09902110, r14922012, b11701201, f09922081, sdlin\}@csie.ntu.edu.tw} \\}

\begin{document}
\maketitle
\begin{abstract}
Conversational recommender systems (CRS) increasingly rely on user simulators for automated evaluation of sales agents. A key requirement for such simulators is the ability to model human decision-making. However, most LLM-based simulators often exhibit unrealistically strong information-processing capabilities, rarely exhibiting the hesitation and decision deferral commonly observed in real consumer behavior, resulting in overly high acceptance rates.
To address this limitation, we propose Hesitator, a theory-grounded user simulation framework that explicitly models human decision-making under choice overload. The framework introduces a modular Decision Module that separates utility-based item selection from overload-aware commitment decisions. Experiments across multiple user simulation frameworks, domains, sales modes, and LLM backbones show that integrating our module consistently mitigates unrealistic behaviors under increasing overload conditions. Furthermore, Hesitator reproduces established behavioral patterns from psychological economics, demonstrating its ability to model human decision behavior.

\end{abstract}
\section{Introduction}
Large language model (LLM) agents have recently been deployed across a wide range of applications, including conversational assistants \cite{achiam2023gpt}, autonomous task planning and tool use \cite{yao2022react, schick2023toolformer}, and complex multi-step reasoning tasks \cite{wei2022chain}. Among these, conversational recommender systems (CRS) represent a particularly compelling use case, where agent-driven dialogues guide users toward purchase decisions. Building effective sales agents for CRS, however, demands reliable automated evaluation pipelines, which in turn depend on realistic user simulation \cite{zhang2020evaluating, yoon2024evaluating}. In such simulated environments, faithfully modeling human decision-making is critical for assessing CRS effectiveness \cite{yoon2024evaluating}. When user simulators fail to capture the gap between stated intentions and actual behavior, estimated purchase probabilities tend to be overly optimistic, obscuring the intention--behavior frictions that reduce real-world conversion rates \cite{sheeran2002intention, carrington2014lost}.

\begin{figure}[t]
  \centering
  \includegraphics[width=\columnwidth]{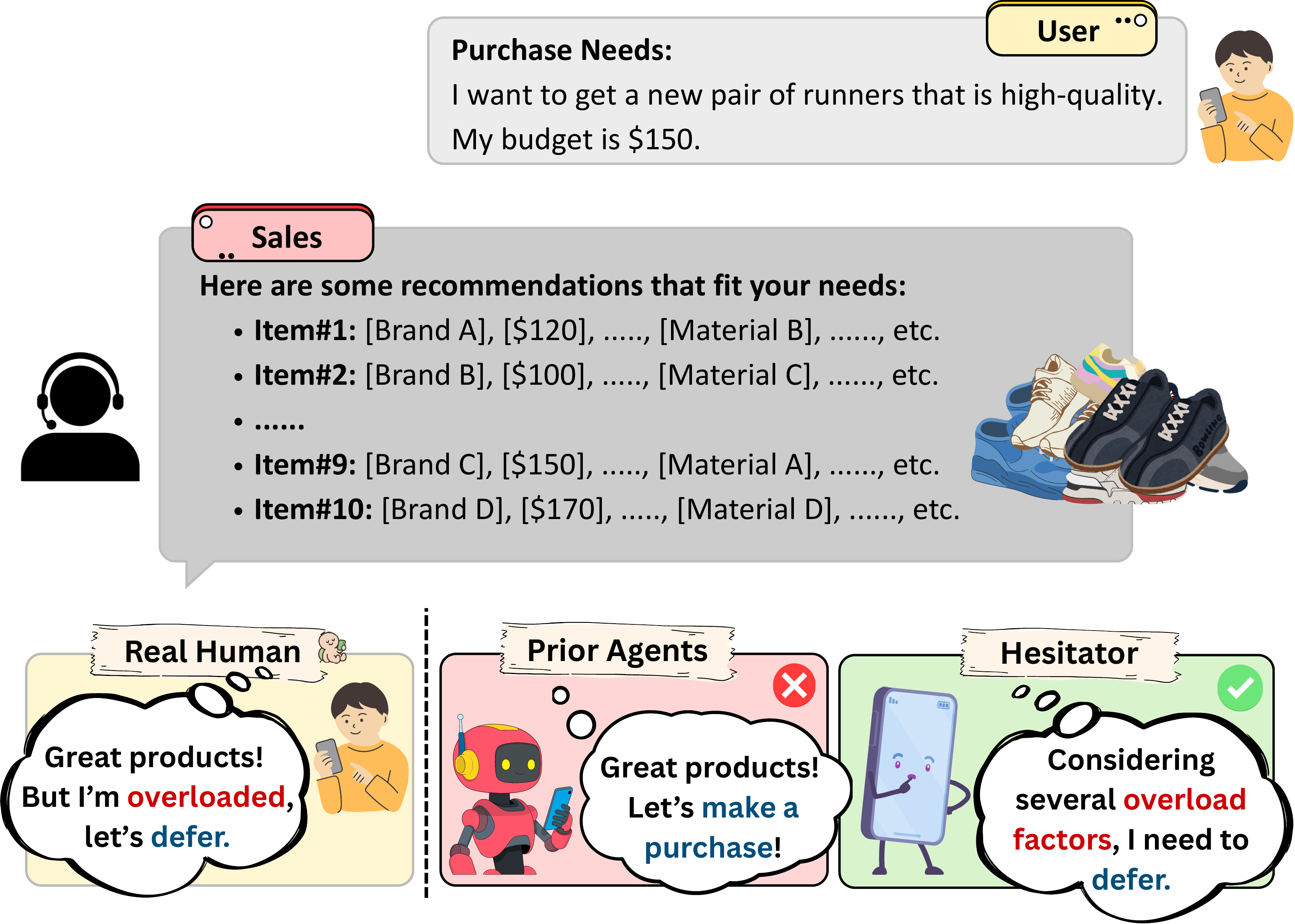}
  \caption{Motivation of Hesitator. Prior user agents often make immediate purchase decisions after identifying good items, ignoring cognitive constraints. In contrast, humans may defer decisions under choice overload. This gap motivates Hesitator, which models hesitation by incorporating overload factors.}
  \label{fig:intro}
\end{figure}
A prominent research direction for evaluating sales agents involves simulating human behavior through LLM-based agents. For instance, RecUserSim implements a multi-module LLM agent architecture with persona, memory, and action modules, whereas UserSimCRS follows an agenda-based framework relying on predefined agenda transitions and predefined goals. While these works have demonstrated strong capabilities in modeling realistic individual personas and diverse user populations~\cite{zhao2025personalens, bernard2025usersimcrs, chen2025recusersim}, a critical dimension remains largely overlooked: the cognitive decision processes underlying user choices~\cite{bettman1998constructive}.
Specifically, in real-world settings, users do not always respond to recommendations immediately or decisively. They may defer decisions when presented with an excessive number of options. This is a well-documented phenomenon known as choice overload. However, because LLM-based user agents inherit the strong information-processing capabilities of their underlying models~\citep{brown2020language, minaee2024large}, they can handle large volumes of unstructured information without exhibiting such cognitive strain. As a result, existing user simulation frameworks tend to underrepresent the hesitation and decision deferral that commonly arise in real consumer behavior.

To address this limitation, we propose \textit{Hesitator}, a decision-aware user simulation framework built around a Decision Module with two cognitively grounded components. The \textbf{Selection Module} employs a two-stage decision process from \citet{bettman1998constructive}, applying non-compensatory filtering to prune the candidate set before compensatory evaluation selects the final item. Without this explicit filtering stage, LLM-based agents tend to give equal consideration to all presented items, which does not reflect how real users selectively attend to options under cognitive constraints. The \textbf{Hesitation Module} quantifies choice overload through a four-dimensional overload vector whose dimensions correspond directly to the moderator categories identified in the meta-analysis of \citet{chernev2015choice}: the number of alternatives, attribute complexity, decision goal, and preference uncertainty. Crucially, rather than relying on LLM judgment alone, we leverage the regression coefficients derived from that meta-analysis to construct a calibrated mapping function that translates the overload vector into an acceptance probability, grounding the agent's deferral behavior in large-scale empirical evidence rather than model priors.

Experimental results yield two key findings. First, \textit{Hesitator} reproduces well-established results from behavioral economics, including the inverted-U effect of total information load~\cite{jacoby1974brand}, the increasing-then-plateau effect of attribute information~\cite{fasolo2007escaping}, and decision conflict theory~\cite{tversky1992choice, anderson2003psychology}. Second, integrating our decision module into three existing user simulation frameworks demonstrates that this lightweight addition significantly mitigates unrealistic user behaviors across diverse datasets, sales agent configurations, and LLM backbones. Notably, while existing user agents exhibit flat or even increasing acceptance rates under severe choice overload, augmented agents with our decision module display a clear decline consistent with human cognitive patterns. These results together demonstrate that \textit{Hesitator} reliably models \hl{the deferral aspect of} human decision behavior \hl{under} diverse overload conditions. We summarize our contributions as follows:

\begin{itemize}[leftmargin=*]
    \item We identify a key limitation of existing LLM-based user simulators for conversational recommender systems (CRS): their lack of cognitive constraints when processing large amounts of information often leads to unrealistically high acceptance rates, inflating estimates of CRS performance.

    \item We propose a plug-and-play \textit{Decision Module} that models intention–behavior friction under choice overload, enabling simulated users to adapt their commitment decisions according to the cognitive load of the decision environment.

    \item We introduce \textit{Hesitator}, a modular user simulation framework that integrates the Decision Module and reproduces established behavioral patterns from decision-making research, enabling more realistic and reliable CRS evaluation.
\end{itemize}
\section{Preliminary}
\label{sec:preliminary}

\subsection{Psychological Economics Theory}
\paragraph{Inverted U Information Overload Curve}

\begin{figure}[ht]
  \includegraphics[width=\columnwidth]{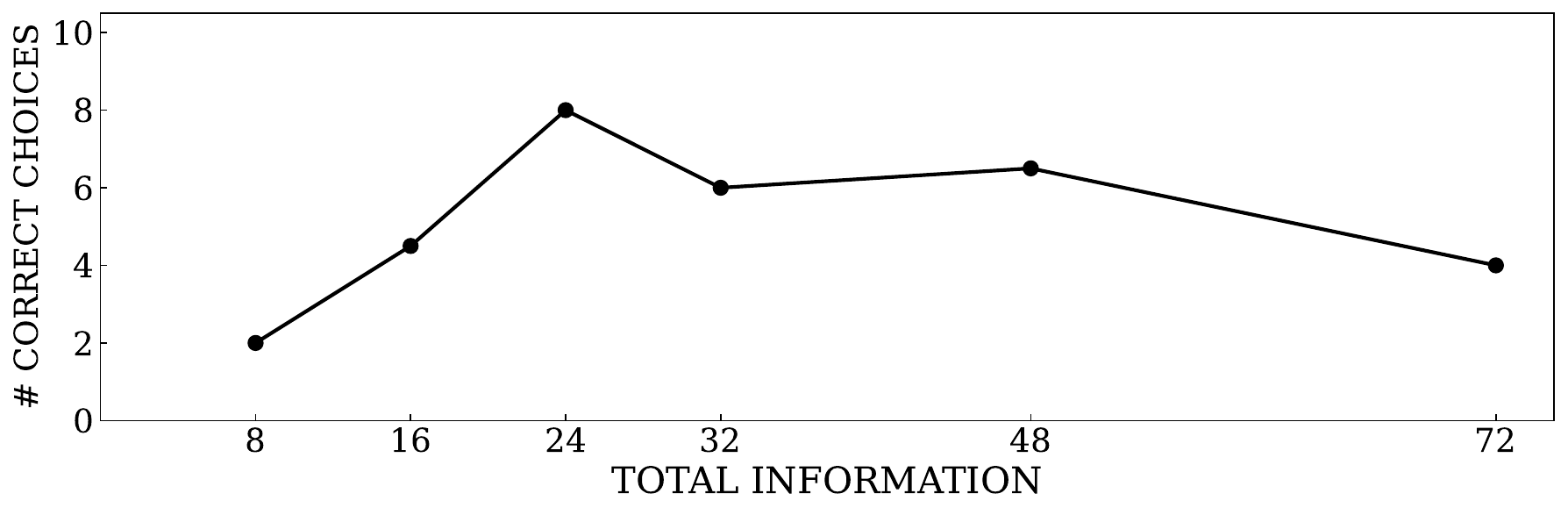}
  \caption{Illustration of the inverted-U relationship between information load and decision quality in consumer decision research \cite{jacoby1974brand}. As information increases, decision performance initially improves but eventually declines due to cognitive overload.}
  \label{fig:jacoby_curve}
\end{figure}
Early work in consumer psychology established the concept of information overload. As shown in Figure~\ref{fig:jacoby_curve}, prior studies proposed an inverted-U relationship between total information load and consumer decision quality \citep{jacoby1974brand}. This model jointly considers assortment size and attribute quantity, suggesting that decision performance improves up to an optimal point but subsequently declines once cognitive load exceeds an individual's processing capacity \citep{eppler2004concept, lurie2004decision}. Furthermore, meta-analytic evidence suggests that individuals with lower preference uncertainty are less susceptible to information overload, as experts are better able to process and filter large amounts of information than novices \citep{chernev2015choice}.

\paragraph{Independent Effects of Assortment and Attributes}
Subsequent research suggests that assortment size and the number of attributes should be treated as independent factors rather than a single aggregate information measure \citep{malhotra1982information, keller1987effects}. Increasing the number of attributes can improve decision performance by providing more diagnostic information \citep{fasolo2007escaping}. In contrast, larger assortments may impair decision performance when additional options introduce substantial tradeoffs. According to decision conflict theory, when alternatives possess complementary attributes such that none clearly dominates, consumers experience psychological conflict \citep{tversky1992choice, anderson2003psychology}, which often leads to decision deferral or reduced transaction likelihood \citep{iyengar2000choice, chernev2003more}.

\subsection{Notations}
\label{sec:prelim_notation}

The notation used throughout this work is defined as follows. Calligraphic letters (e.g., $\mathcal{G}$) denote high-level modules or sets, bold symbols (e.g., $\mathbf{v}$) represent multidimensional vectors, and Greek letters (e.g., $\phi, \sigma$) indicate internal persona traits or external scenario constraints. Table~\ref{tab:notations} summarizes the main symbols.

\paragraph{Agent State Representation}
The core of our framework is the global user agent state $\mathcal{G} = (\mathcal{P}, \mathcal{S})$, which encapsulates the interaction between a user's stable characteristics and their immediate environment. The \textbf{persona} ($\mathcal{P}$) represents the intrinsic psychological profile of the user, parameterized by a vector of traits $\boldsymbol{\phi} = [\phi_O, \phi_K, \phi_U]^\top$ corresponding to \textit{Openness}, \textit{Pickiness}, and \textit{Preference Uncertainty}. The \textbf{scenario} ($\mathcal{S}$) captures the contextual conditions of the recommendation task through environmental constraints $\boldsymbol{\sigma} = [\sigma_N, \sigma_B, \sigma_T]^\top$, representing \textit{Current Needs}, \textit{Budget}, and \textit{Time Pressure}. In addition, the \textbf{dialogue history} ($\mathcal{H}$) records the sequential interaction between the agents, formulated as $\mathcal{H} = \{(u_0), (r_1, u_1), \dots, (r_t, u_t)\}$, where $r$ denotes the response of the Sales Agent and $u$ denotes the response of the User Agent.



\paragraph{Cognitive Overload Hierarchy}
\label{sec:prelim_overload}

Following the theoretical framework of \citet{chernev2015choice}, the consumer cognitive state is modeled using a multidimensional overload vector $\mathbf{v} = [v_a, v_s, v_t, v_u]^\top$. This vector decomposes the choice environment into four primary antecedents: \textbf{Assortment Size} ($v_a$), \textbf{Choice Set Complexity} ($v_s$), \textbf{Task Difficulty} ($v_t$), and \textbf{Preference Uncertainty} ($v_u$). These factors are further organized into a hierarchical structure of leaf variables. 
The aggregate cognitive impact of these factors is quantified as the total overload effect size $d_{\text{total}}$, which serves as the primary latent variable governing the agent's decision to accept or reject.

\subsection{Simulation Process}

The simulation framework operates in a CRS as a turn-based dialogue between a Sales Agent and a User Agent. The interaction is governed by a global user state $\mathcal{G}=(\mathcal{P},\mathcal{S})$, where $\mathcal{P}$ denotes the user persona and $\mathcal{S}$ the shopping scenario. A dialogue history $\mathcal{H}=\{(u_0),(r_1,u_1),\dots,(r_t,u_t)\}$ is maintained, where $r_t$ and $u_t$ denote the responses of the Sales Agent and User Agent at turn $t$, respectively.

The interaction is initialized by the User Agent with an initial message $u_0$ describing the user's needs given $\mathcal{G}$. At each turn $t$, the Sales Agent generates $r_t$ by retrieving candidate items from the database based on inferred preferences and dialogue history $\mathcal{H}$. The User Agent then evaluates $r_t$ together with $\mathcal{G}$ and produces $u_t$, which includes the decision outcome (e.g., accept, reject, or defer) and an explanation when rejecting the recommendation. The dialogue history is updated as $\mathcal{H}\leftarrow\mathcal{H}\cup\{(r_t,u_t)\}$, and the interaction continues until a purchase decision is reported or the predefined turn limit $T$ is reached.

\section{Method}
\label{sec:hesitator}

\begin{figure*}[t]
  \centering
  \includegraphics[width=\textwidth]{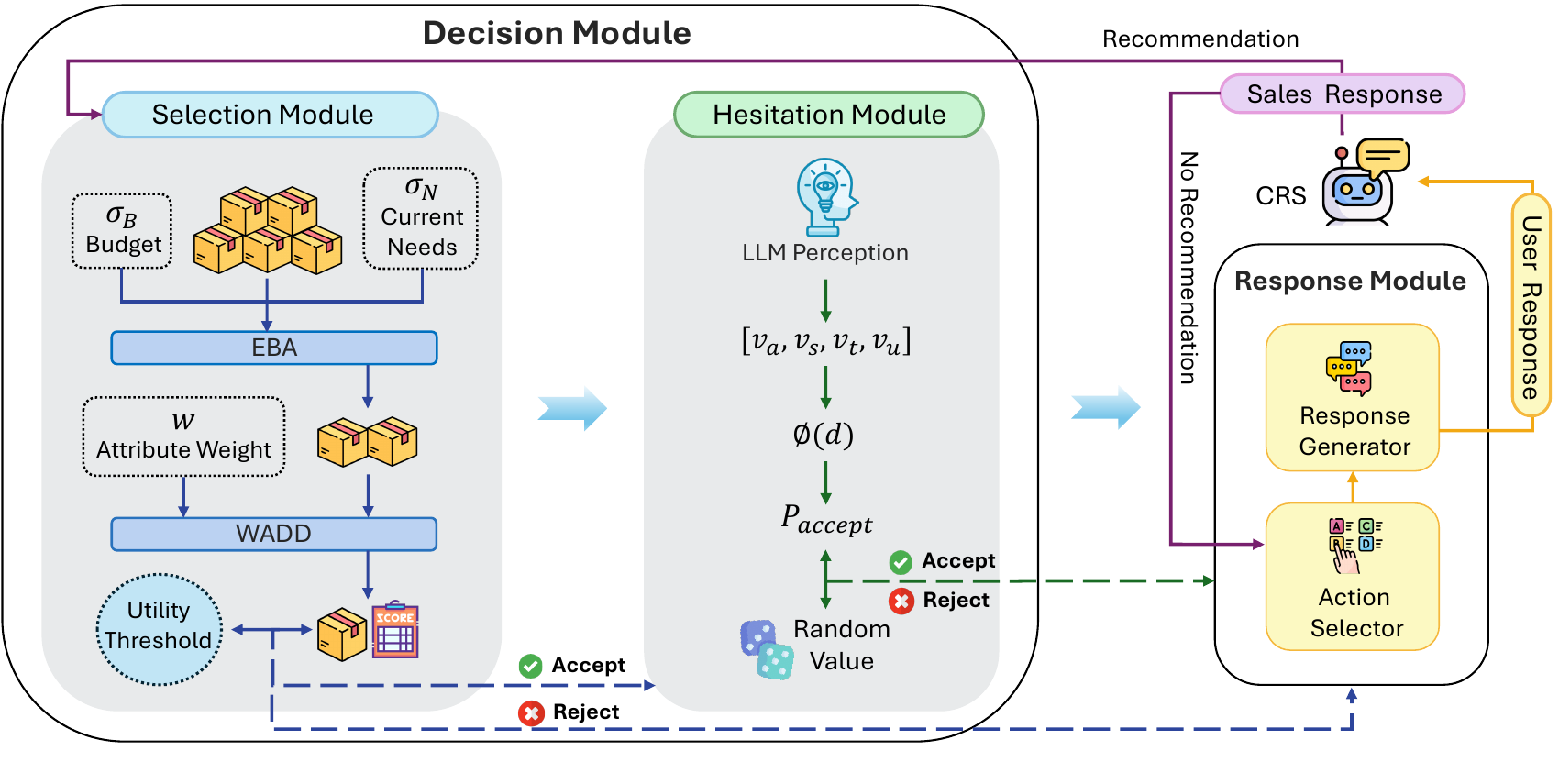}
  \caption{Architecture of the Hesitator framework. The Selection Module filters and ranks items (EBA → WADD). If the recommendation passes the selection stage (accept), it proceeds to the Hesitation Module, which estimates the final acceptance probability based on perceived overload factors; otherwise, the system directly generates a rejection response. Solid lines indicate data flow, while dashed lines denote internal stage transitions within the user agent.}
  \label{fig:hesitator}
\end{figure*}
In this section, we present \textit{Hesitator}, an LLM-based framework that simulates user decisions via three different modules.
As illustrated in Fig.~\ref{fig:hesitator}, the Decision Module consists of a \textit{Selection Module} that identifies items satisfying the user's needs and a \textit{Hesitation Module} that models commitment decisions under cognitive overload. To support conversational recommendation, a \textit{Response Module} further converts internal decisions into natural language responses.


\subsection{Decision Module}
\label{sec:decision}
Real-world decisions involve both evaluating option quality and determining whether to commit under cognitive constraints. Accordingly, the \textit{Decision Module} is formulated as a conditional sequential pipeline: first, the \textit{Selection Module} evaluates the intrinsic utility of the recommended items $\mathcal{I}$ based on the last sales agent recommendation response $r_t$. If the options fail to meet basic expectations (encoded in $\mathcal{G}$), the agent directly rejects the recommendation at this turn. Conversely, if an acceptable option is identified, the process advances to the \textit{Hesitation Module}, which models choice overload to determine whether the user commits to the choice or defers. The final outcome $\mathcal{D}$ is subsequently routed to the \textit{Response Module}.

\subsubsection{Selection Module}
\label{subsec:selection}
Evaluating multiple recommended items can be cognitively demanding, making it impractical to exhaustively compare all options. To approximate human decision strategies, the \textit{Selection Module} adopts a two-stage heuristic-to-compensatory process, which is commonly observed in human decision-making \citep{bettman1998constructive}. 
\hl{This pipeline is grounded in the constructive consumer choice framework of} \citet{bettman1998constructive} \hl{and represents one cognitively plausible family of heuristics that fits attribute-driven shopping behavior---the dominant regime in CRS research, where users express constraints and preferences over discrete attribute dimensions such as price, brand, and specifications. We do not claim it as a universal model of human decision-making; less attribute-driven settings (e.g., aesthetic-driven, emotion-driven, or exploratory shopping) may follow different cognitive strategies and motivate alternative selection strategies as future work.}
Specifically, it first eliminates options that violate the user's hard constraints and then estimates compensatory utility over the remaining candidates.

\noindent\textbf{Stage 1: Non-compensatory Filtering.} 
If the initial set $|\mathcal{I}| > \theta$, the agent employs Elimination by Aspects (EBA). An item $i \in \mathcal{I}$ is retained in the candidate set $\mathcal{C}$ if it satisfies all critical constraints $\mathcal{K}$ (e.g., price $\leq \sigma_B$):
\begin{equation}
    \mathcal{C} = \{i \in \mathcal{I} \mid \forall k \in \mathcal{K}, \mathbf{I}_k(i) = 1\}
\end{equation}

\noindent\textbf{Stage 2: Compensatory Evaluation.} 
When $|\mathcal{C}| \leq \theta$ (we set $\theta = 3$ following established heuristics~\cite{bettman1998constructive}), the agent evaluates utility $u_i$ via a Weighted Additive (WADD) function:
\begin{equation}
    u_{i} = \mathbf{w}^\top \mathbf{a}_i = \sum_{j=1}^{n} w_{j} \cdot a_{ij}
\end{equation}
where $a_{ij}$ is the attribute vector perceived by the LLM and $w_{j}$ is extracted from historical reviews (see Appendix~\ref{app:user_profile}).
\paragraph{Acceptance Mechanism} 
Following compensatory evaluation, the selection outcome $s=\{s_u,s_i\}$ is defined by the highest utility $s_u=\max_{i\in\mathcal{C}} u_i$ and the corresponding item $s_i=\arg\max_{i\in\mathcal{C}} u_i$. The agent compares $s_u$ with a pickiness-dependent threshold $\tau(\phi_K)=\gamma+\alpha\phi_K$ (with $\gamma=0.6,\ \alpha=0.1$), where $\gamma$ and $\alpha$ are tunable hyperparameters used to simulate different user acceptance levels. If $s_u<\tau(\phi_K)$, the recommendation is rejected and passed to the \textit{Response Module}; otherwise, it proceeds to the \textit{Hesitation Module} $f_{\text{hesitate}}$ to evaluate potential deferral.
\hl{The Selection Module is architecturally modular: alternative selection strategies 
can be substituted as a drop-in replacement without modifying the Hesitation Module or Response Module.}

\subsubsection{Hesitation Module}
\label{subsec:hesitation}
Even when a satisfactory option exists, users may defer decisions due to cognitive overload \citep{chernev2015choice}. To capture this phenomenon, the \textit{Hesitation Module} models the probability that a user commits to the identified option given the perceived cognitive load of the decision environment. This module operates in three phases: (1) inferring the user’s cognitive state ($\mathbf{v}$) from dialogue, (2) calibrating these observations to behavioral effect sizes ($d_\text{total}$), and (3) mapping the aggregated overload to the final acceptance probability ($P_{\text{accept}}$).

\noindent\textbf{Phase 1: State Perception.}
The framework elicits granular intensity scores $v \in \{1, 2, 3\}$ for five leaf dimensions (i.e., $v_a, v_{s,d}, v_{s,a}, v_{t,a}$, and $v_{t,f}$) via LLM-based prediction. The remaining leaf variables, $v_u$ and $v_{t,p}$, are predefined constants determined by the user persona ($\phi_U$) and task constraints ($\sigma_T$). Composite factors $v_k \in \{v_s, v_t\}$ are formally defined as the arithmetic mean of their respective constituent leaf sets.

\noindent\textbf{Phase 2: Meta-Analytic Calibration.}
\hl{The Hesitation Module needs to map a perceived overload state into a real-valued acceptance probability, and this mapping should be grounded in evidence about real human deferral behavior rather than model priors. Relying on an LLM alone to judge how much a given overload level should reduce commitment willingness would require fine-grained behavioral calibration that LLMs handle poorly and inconsistently. Instead, we leverage the meta-analytic regression of} \citet{chernev2015choice}\hl{, which aggregates effect sizes across 99 published experimental observations and provides an empirically estimated mapping from overload moderators to standardized effect sizes (Cohen's $d$). By inheriting their regression coefficients, our module obtains a calibration that reflects population-scale human decision behavior.}
To connect discrete LLM-based perceptions with continuous behavioral effects, we map perceived overload factors to empirical effect sizes derived from the meta-analysis of \citet{chernev2015choice}. This meta-analysis reports standardized effect sizes (Cohen’s $d$) across heterogeneous experimental settings.
\hl{Specifically, $\delta$ encodes, for each overload factor, the minimum and maximum effect sizes observed across the meta-analyzed studies. We use $\delta$ to construct the interpolation function $f_{\text{interp}}(\cdot)$, which maps a discrete LLM-perceived overload level $v \in \{1,2,3\}$ onto a continuous effect-size value within this empirically observed range. Without $\delta$, the perceived overload levels would have no behavioral scale; $\delta$ is what gives them magnitude grounded in empirical evidence.}

Four key overload factors are perceived by an LLM: \textbf{assortment size} ($v_a$), \textbf{choice set complexity} ($v_s$), \textbf{task difficulty} ($v_t$), and \textbf{preference uncertainty} ($v_u$), while factors with negligible effects in CRS are omitted (Appendix~\ref{subsec:factors_omitted_rationale}). Each perceived factor level is mapped to an empirical effect-size range via $f_{\text{interp}}(\cdot)$.

Following the meta-analytic regression formulation, the aggregate overload effect is computed as:
\begin{equation}
d_{total} =
\boldsymbol{\beta}^{\top}
f_{\text{interp}}(\mathbf{v}, \delta)
\end{equation}
where $\mathbf{v} = [v_a, v_s, v_t, v_u]$ denotes the perceived factor levels and $\boldsymbol{\beta}$ are moderator coefficients from the meta-analysis. Detailed mapping procedures and parameter settings are provided in Appendix~\ref{subsec:regression_model}.

\noindent\textbf{Phase 3: Probabilistic Mapping.}
To convert the aggregated overload effect size into a decision probability, we adopt the arcsine-based proportion transformation commonly used in statistical power analysis \citep{cohen2013statistical,fleiss2013statistical}. Rearranging the formulation yields the following closed-form expression:

\begin{equation}
\label{eq:p_accept}
P_{\text{accept}} = \sin^2\left(\arcsin(\sqrt{P_{\text{base}}}) - \frac{d_{\text{total}}}{2}\right)
\end{equation} 

The derivation is provided in Appendix~\ref{subsec:d_value_derivation}. Here, $P_{\text{base}} = 0.5$ represents a neutral prior that balances the bidirectional effects of the “more-is-better” tendency and choice overload (see Appendix~\ref{subsec:p_base_design}). The value of $d_{total}$ is bounded to ensure that the resulting probability remains within $[0,1]$. With $P_{\text{base}}=0.5$, $d_{total}$ is truncated to the range $[-\pi/2,\pi/2]$. A purchase is triggered only if a random sample $\epsilon \leq P_{\text{accept}}$; otherwise, the agent defers the decision and provides a rationale.


\subsection{Response Module}
\label{subsec:response}
The \textit{Response Module} converts internal decisions into communicative actions \hl{through two prompt-based substeps}. It first performs \textbf{Action Selection} $\mathcal{A} = \text{LLM}(\mathcal{G}, \mathcal{H}, \mathcal{D})$, \hl{prompting an LLM to choose} a high-level communicative intent \hl{(e.g., accept, ask follow-up, request clarification, or defer) given the global state, dialogue history, and decision outcome}. Finally, \textbf{Response Synthesis} $\mathcal{R} = \text{LLM}(\mathcal{G}, \mathcal{H}, \mathcal{D}, \mathcal{A})$ \hl{prompts the LLM to verbalize} the final response $R$ \hl{conditioned on the same context together with the selected action. The full prompts used for both substeps are provided in} Appendix~\ref{app:response_module_prompts}.

\section{Experiments}
\label{sec:experiments}

\subsection{Experimental Setup}
\label{subsec:setup}

\begin{figure*}[ht]
  \centering
  \includegraphics[width=\textwidth]{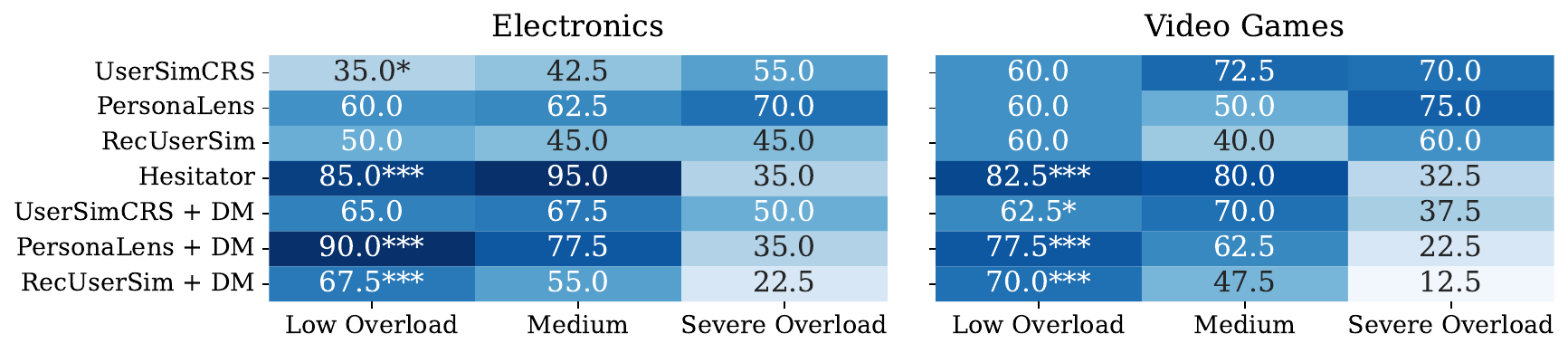}
  \caption{Simulation success rates (\%) of different user agents under varying levels of cognitive overload in two domains (Electronics and Video Games). The Decision Module (+DM) is designed to mitigate the original tendency of user agents to remain insensitive to cognitive overload, enabling their decision behavior to vary more appropriately with overload level. As a result, success rates decline under severe overload, better reflecting human decision patterns. Statistical significance is determined using the Wilcoxon signed-rank test(Low vs. Severe overload); * ($p < 0.05$), ** ($p < 0.01$), and *** ($p < 0.001$).}
  \label{fig:basic_overload}
\end{figure*}

\noindent\textbf{Simulation Environment.}
We simulate conversational recommendation interactions using two domains from the \textbf{Amazon Reviews 2023} dataset \cite{hou2024bridging}: \textit{Electronics} and \textit{Video Games}. Each experimental configuration is evaluated over $N=40$ independent sessions. A session follows a turn-based dialogue history $\mathcal{H}$ and imposes a maximum of $T=20$ interaction turns. Each session is initialized with a unique global state $\mathcal{G}$ that defines the user persona and shopping scenario.

\noindent\textbf{User Profiles.}
The user state $\mathcal{G}$ consists of a persona $\mathcal{P}=\{\phi_O,\phi_K,\phi_U\}$ and a shopping scenario $\mathcal{S}=\{\sigma_N,\sigma_B,\sigma_T\}$. Persona traits include openness, pickiness, and preference uncertainty, while the scenario specifies the user’s needs, budget, and time pressure, without revealing the item title to avoid data leakage \cite{zhu2024reliable}.

\noindent\textbf{Sales Agent.}
We adopt CSI \cite{kim2025towards} as the sales agent framework and use \texttt{gpt-oss-20b} as the backbone model. The agent retrieves candidate items through a vector retrieval system built on \textit{Qwen3-Embedding-0.6B}. It operates in two modes: \textbf{Basic} (preference probing and suggestion) and \textbf{Persuasive} (strategic persuasion actions). 

\noindent\textbf{Baselines.}
We compare our method against three representative user simulators: \textbf{PersonaLens} \cite{zhao2025personalens}, \textbf{UserSimCRS} \cite{bernard2025usersimcrs}, and \textbf{RecUserSim} \cite{chen2025recusersim}, the current SOTA user simulator for CRS. All baselines share the same backbone model (\texttt{gpt-oss-20b}) and identical initialization states to ensure a controlled comparison.

\noindent\textbf{Evaluation Metrics.}
We evaluate the user agent systems using three complementary metrics: \textbf{Success Rate (SR)}, \textbf{Subjective Dialogue Quality} \cite{chen2025recusersim}, and \textbf{Believability} \cite{xiao2023far}. SR measures the proportion of sessions ending with a purchase decision. Dialogue quality is assessed through pairwise comparisons along five dimensions (realism, naturalness, relevance, clarity, adaptability). Believability is measured via the hallucination ratio $\bar{H}_R$. 
Detailed system implementation and metrics are provided in Appendix~\ref{app:implementation} and ~\ref{app:exp_details}, respectively.

\begin{figure*}[t]
\centering

\begin{subfigure}{0.32\textwidth}
  \centering
  \includegraphics[width=\linewidth]{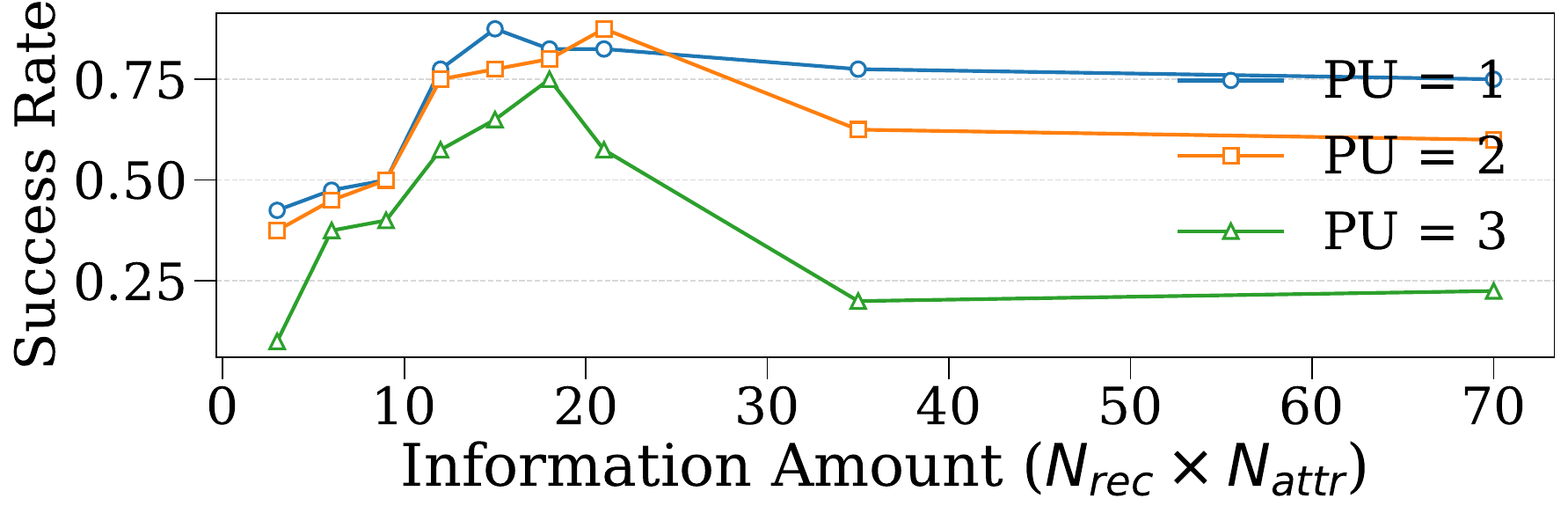}
  \caption{Success rate vs. total information}
  \label{fig:total_info}
\end{subfigure}
\hfill
\begin{subfigure}{0.32\textwidth}
  \centering
  \includegraphics[width=\linewidth]{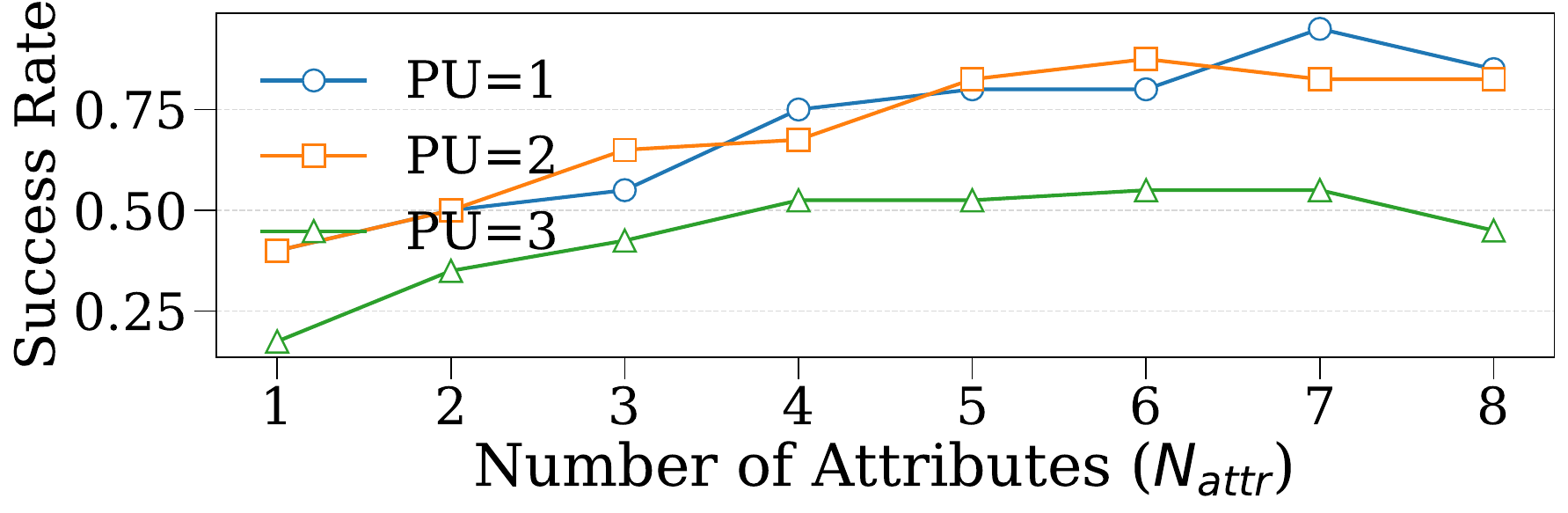}
  \caption{Success rate vs. number of attributes}
  \label{fig:na}
\end{subfigure}
\hfill
\begin{subfigure}{0.32\textwidth}
  \centering
  \includegraphics[width=\linewidth]{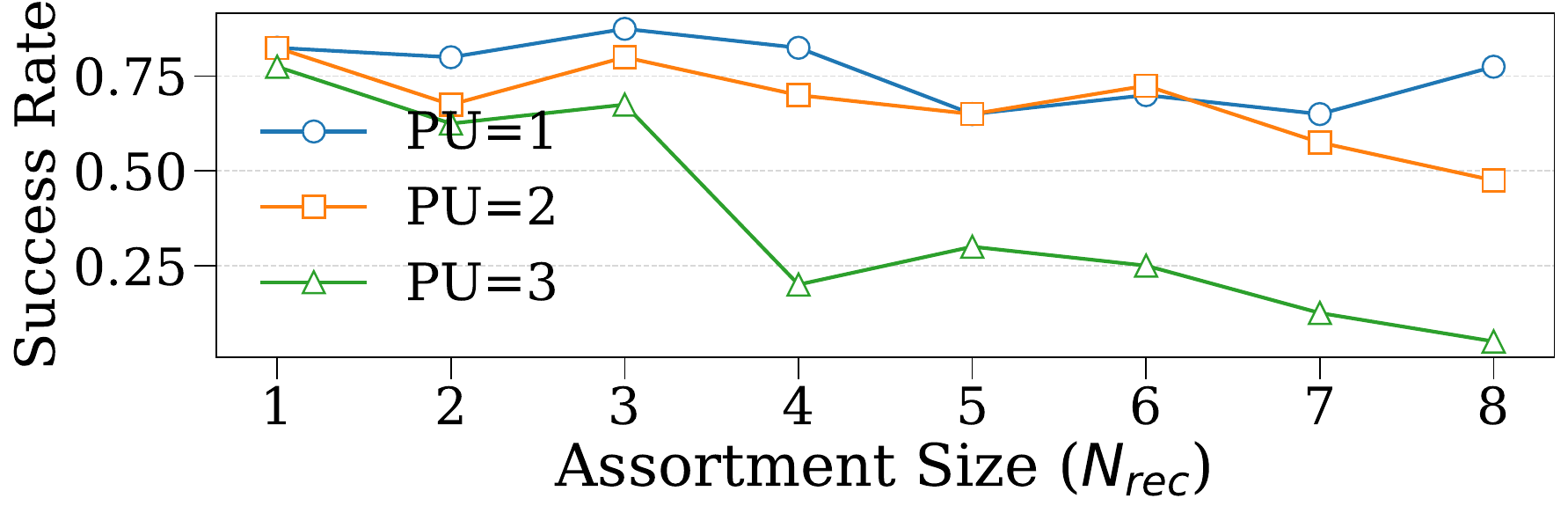}
  \caption{Success rate vs. assortment size}
  \label{fig:as}
\end{subfigure}

\caption{
Decision success under different information conditions and preference uncertainty (PU). 
(a) Total information exhibits an inverted-U relationship with success rate, consistent with information overload findings. 
(b) Increasing attribute information improves success initially but shows diminishing returns. 
(c) Larger assortments reduce success under high uncertainty, reflecting decision conflict and choice overload.
}
\label{fig:info_overload}
\end{figure*}


\subsection{Overload Effect}
\label{subsec:overload_exp}

We analyze how user-agent decision behavior changes under different cognitive overload levels. Specifically, we evaluate three conditions: \textit{Low}, \textit{Medium}, and \textit{Severe}. The exact parameter settings for these conditions are described in Appendix~\ref{app:overload_config}.

\paragraph{Anomalous Baseline Behaviors}

Figure~\ref{fig:basic_overload} shows the success rates of different user-agent frameworks under increasing cognitive overload. Without the Decision Module, baseline simulators exhibit unrealistic patterns: their success rates remain stable or even increase as overload intensifies from Low to Severe. For example, in the Electronics domain, \textit{PersonaLens} rises from 60.0 to 70.0 and \textit{UserSimCRS} from 35.0 to 55.0. Similar trends appear in the Video Games domain, where \textit{PersonaLens} reaches 75.0 and \textit{UserSimCRS} maintains 70.0 under Severe overload. These behaviors contradict well-established findings in consumer decision-making, where increasing overload typically raises cognitive conflict and leads to decision deferral.

\paragraph{Behavioral Correction via Decision Module}

After integrating the Decision Module (+DM), all baseline user simulators exhibit behavior consistent with human decision patterns: success rates decline as overload increases from Low to Severe. For example, in the Electronics domain, \textit{RecUserSim+DM} drops from 67.5 to 22.5 and \textit{PersonaLens+DM} from 90.0 to 35.0. A similar trend is observed for \textit{UserSimCRS+DM}, which decreases from 65.0 to 50.0. Consistent patterns also appear in the Video Games domain, where \textit{RecUserSim+DM} falls from 70.0 to 12.5 and \textit{PersonaLens+DM} from 77.5 to 22.5 under Severe overload. Overall, these results indicate that the Decision Module corrects unrealistic decision behavior in existing simulators and enables them to reproduce the decline in decision commitment commonly observed under severe cognitive overload. Additional results across different sales agents and LLM backbones are provided in Appendix~\ref{app:exp_var}.

\subsection{Analysis on Psychological Economics Theory}
\label{subsec:psycho_repro}
To further evaluate the realism of the proposed user agent, we examine whether \textit{Hesitator} can reproduce established findings in psychological economics. 

\paragraph{Inverted-U Information Overload Curve}
As shown in Figure~\ref{fig:total_info}, the pattern closely follows the classical findings of \citet{jacoby1974brand} (Figure~\ref{fig:jacoby_curve}): success rates initially increase as information grows, peak at moderate information levels, and then decline once cognitive load becomes excessive, forming an inverted-U relationship between information quantity and decision performance.
Notably, the decline becomes less pronounced as preference uncertainty decreases. Agents with lower uncertainty exhibit a flatter post-peak slope, indicating greater resilience to information overload. This reproduces a well-documented behavioral phenomenon in consumer decision-making: individuals with higher expertise are less likely to defer decisions under high information load.

\paragraph{Independent Effects of Assortment and Attributes}
We further analyze the independent effects of assortment size ($N_{rec}$) and attribute quantity ($N_{attr}$), as illustrated in Fig.~\ref{fig:na} and Fig.~\ref{fig:as}. Consistent with prior studies \citep{malhotra1982information,keller1987effects}, the two factors exhibit distinct impacts on decision behavior. Increasing attribute information generally improves decision performance by providing more diagnostic cues for evaluating alternatives. In contrast, larger assortments introduce decision conflict among competing options, causing success rates to decline as the number of recommendations increases \citep{tversky1992choice}.


\subsection{Comprehensive Conversation Simulation Evaluation}

\paragraph{Subjective Evaluation}

\begin{figure}[t]
  \centering
  \includegraphics[width=0.8\columnwidth]{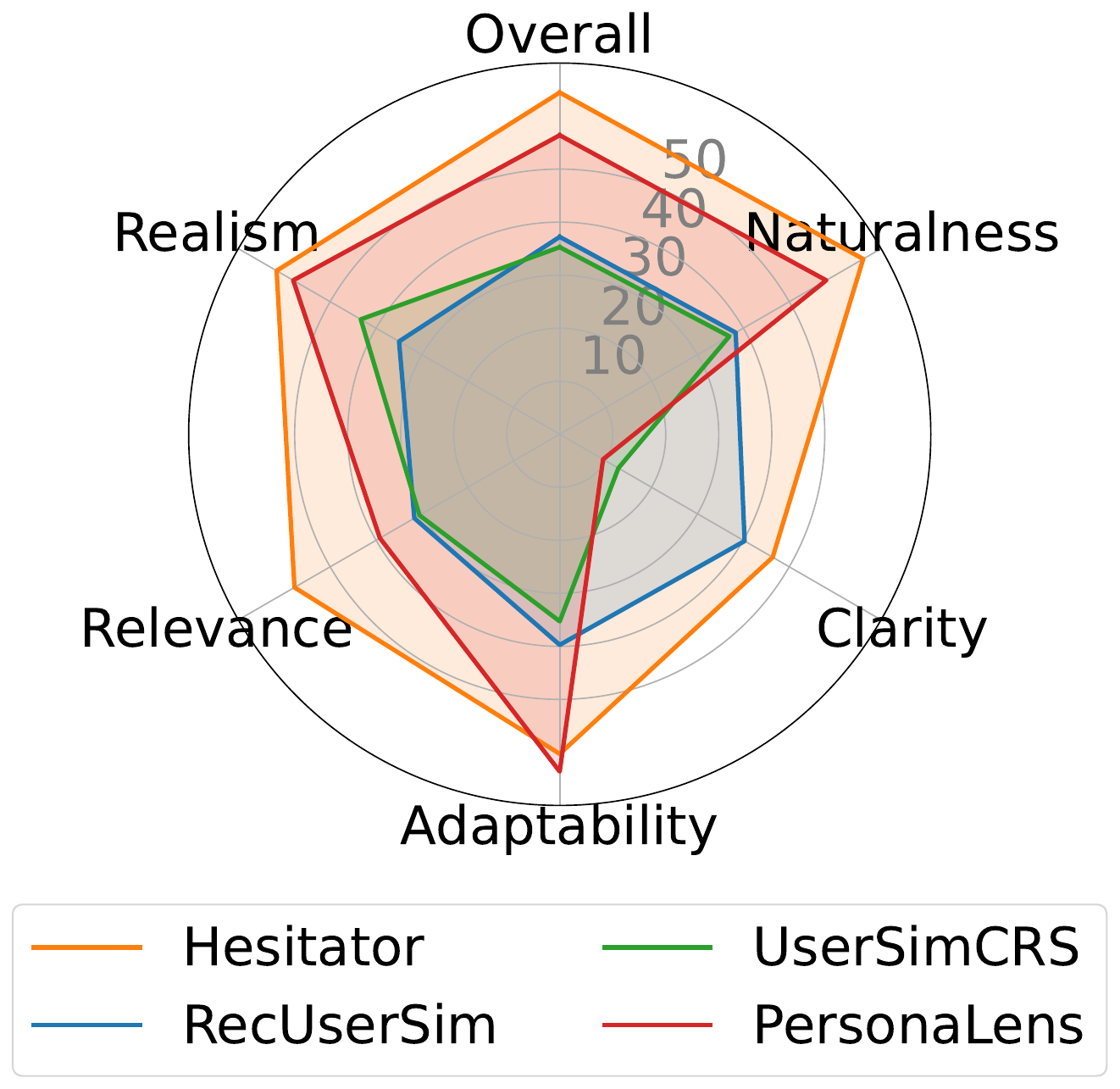}
  \caption{Subjective evaluation of conversational simulation quality across user agents. 
  }
  \label{fig:radar}
\end{figure}

Figure~\ref{fig:radar} compares the conversational simulation quality across user agents. Hesitator consistently outperforms the baseline simulators (RecUserSim, UserSimCRS, and PersonaLens) across multiple evaluation dimensions, including overall quality, naturalness, clarity, adaptability, relevance, and realism. This indicates that our user agent not only models decision behavior more realistically but also maintains strong conversational quality, producing responses that are more natural, coherent, and contextually appropriate than existing simulators. Detailed evaluation setup is provided in Appendix~\ref{app:subjective_metric}

\paragraph{Believability Evaluation}

Table~\ref{tab:hallucination_ratio} reports hallucination ratios under vanilla and decision-module settings. Adding the Decision Module does not increase hallucination and slightly reduces hallucination for most simulators. PersonaLens decreases from 0.1290 to 0.1088 and UserSimCRS from 0.2548 to 0.2268, while RecUserSim changes only marginally from 0.0584 to 0.0624. These results suggest that the Decision Module preserves response believability without introducing additional hallucination. Hesitator nevertheless exhibits a higher hallucination ratio than RecUserSim, likely because the latter includes a dedicated response refinement stage that helps reduce hallucinations before output.
\begin{table}[t]
\centering
\begin{tabular}{lcc}
\toprule
\textbf{Model} & \textbf{Vanilla} & \textbf{DM} \\
\midrule
Hesitator   & 0.0969 & --       \\
PersonaLens & 0.1290 & 0.1088  \\
RecUserSim  & 0.0584 & 0.0624  \\
UserSimCRS  & 0.2548 & 0.2268  \\
\bottomrule
\end{tabular}
\caption{Average hallucination ratio under vanilla and decision-module (+DM) settings. The Decision Module does not increase hallucination and slightly reduces it for most simulators, indicating preserved response believability.}
\label{tab:hallucination_ratio}
\end{table}


\subsection{Ablation Study}
\begin{figure}[t]
  \centering
  \includegraphics[width=0.48\textwidth]{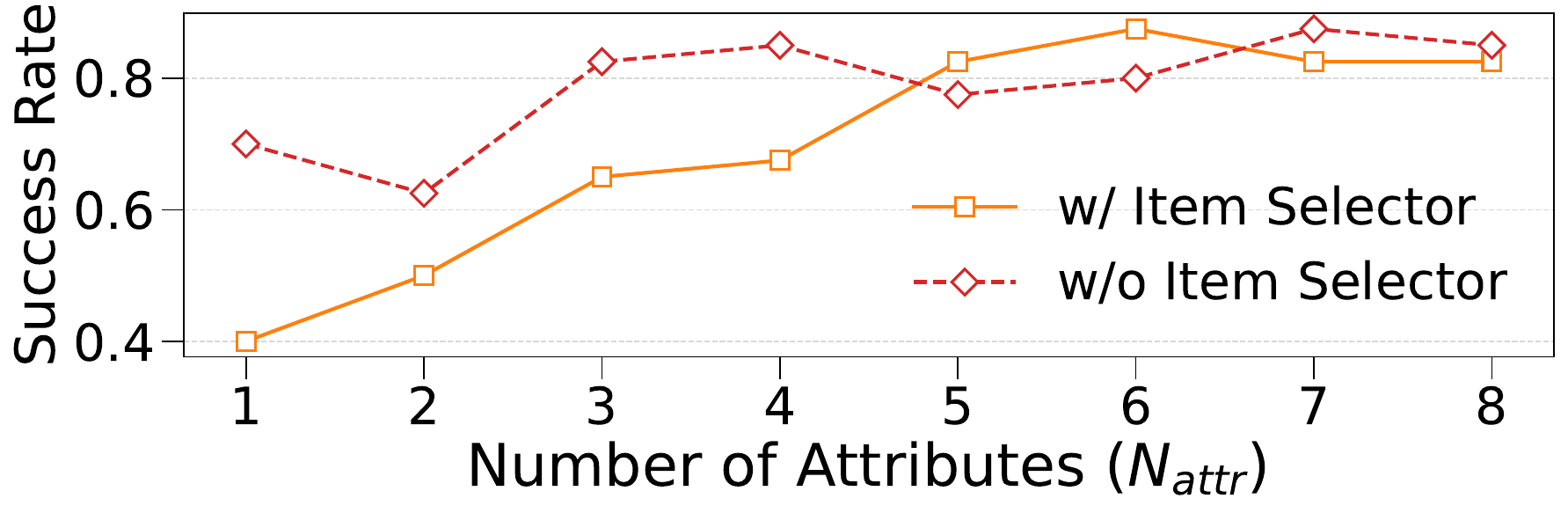}
  \caption{Ablation results for the Selection Module with varying numbers of attributes.}
  \label{fig:abl_sel}
\end{figure}

To isolate the effect of structured item evaluation in the proposed framework, we conduct an ablation study on the \textit{Selection Module} by replacing it with a standard rating-based prompt. The Selection Module models human-like item evaluation via a two-stage decision strategy. As shown in Figure~\ref{fig:abl_sel}, removing this module leads to unstable performance as the number of attributes increases, and the expected positive effect of additional attributes is no longer consistently observed, deviating from human behavior. This suggests that, without structured utility computation, the user agent fails to faithfully reproduce human decision-making.

\section{Related Work}
\paragraph{User Simulation for Conversational Recommendation}

User simulation is widely used to evaluate CRS without requiring costly human studies. Early work relied on rule-based simulators that generate responses using predefined policies or slot-filling strategies \cite{zhang2020evaluating}, offering strong controllability but limited linguistic diversity; later agenda-based simulators model users as goal-driven agents maintaining structured dialogue goals, as exemplified by \textit{UserSimCRS} \cite{bernard2025usersimcrs}. Recent research increasingly adopts LLM-based simulators to produce more natural and context-aware responses. For instance, \textit{PersonaLens} \cite{zhao2025personalens} conditions LLMs on persona descriptions and dialogue context, while \textit{RecUserSim} \cite{chen2025recusersim} introduces a modular architecture to improve behavioral diversity and extensibility. Other work enhances simulation realism by modeling richer user characteristics and evolving intents, such as personality-driven simulators that condition LLMs on stable persona traits \cite{ma2025pub,gromada2025evaluating} and CRS methods that track intent evolution during dialogue (e.g., DICR \cite{zhou2022aligning} and Chat-REC \cite{gao2023chat}). However, LLM-based simulators rely on implicit reasoning without explicitly modeling human decision processes, often producing overly rational behaviors that overlook hesitation under choice overload. We address this by introducing an explicit Decision Module that separates utility-based selection from overload-aware hesitation to enable structured and controllable commitment decisions.

\paragraph{Cognitive Architectures for Decision Modeling}

Cognitive science has long studied computational models of human decision-making. Architectures such as ACT-R \citep{anderson2004integrated} and SOAR \citep{laird2019soar} simulate human cognition through structured modules for perception, memory, and action selection, enabling the modeling of complex behavioral patterns. However, these architectures are typically designed for controlled laboratory tasks and rely on symbolic representations that are difficult to integrate with modern conversational systems. In contrast, our approach augments LLM-based simulators with a theory-grounded Decision Module that models cognitive constraints such as choice overload.


\section{Conclusion}
In this paper, we introduce Hesitator, a theory-grounded user simulation framework designed to model \hl{one important dimension of} human decision-making under cognitive constraints in conversational recommendation\hl{: decision deferral under choice overload}. By integrating a modular Decision Module, composed of a Selection Module that mimics heuristic to compensatory item evaluation and a Hesitation Module that models choice overload, the framework enables LLM-based user agents to simulate both utility-driven selection and overload-induced decision deferral. Experimental results show that incorporating our module consistently corrects unrealistic behaviors in existing simulators under increasing overload conditions while maintaining strong conversational quality. More importantly, Hesitator reproduces established behavioral patterns from psychological economics, demonstrating its effectiveness in \hl{capturing this overload-driven aspect of human-like decision dynamics} and providing a more reliable benchmark for evaluating conversational recommendation systems.

\section*{Limitations}
This work has several limitations. First, our experiments focus on two product domains from the Amazon Reviews dataset. Although these domains are widely used in conversational recommendation research, future work should examine whether similar behavioral patterns emerge in other domains with different decision characteristics. Second, the hesitation mechanism relies on parameters derived from a large-scale meta-analysis in consumer psychology. While this grounding provides empirical support for the modeling assumptions, future work could explore adaptive calibration using domain-specific behavioral data. Finally, user personas are treated as static throughout each interaction session, whereas real-world user preferences may evolve during conversations. Extending the framework to model dynamic preference shifts could further improve simulation realism.


\bibliography{custom}

\appendix
\section{Notation Summary}
\label{app:notations}
\begin{table}[ht]
\centering
\small
\begin{tabularx}{\columnwidth}{@{} l X @{}}
\toprule 
\textbf{Symbol} & \textbf{Description} \\
\midrule
\multicolumn{2}{l}{\textit{State Variables}} \\
\addlinespace[0.5ex]
$\mathcal{G}$ & Global user agent state $(\mathcal{P}, \mathcal{S})$ \\
$\mathcal{P}$ & Persona \\
\quad$\phi_O, \phi_K, \phi_U$ & Openness, Pickiness, and Preference Uncertainty \\
$\mathcal{S}$ & Purchase Scenario \\
\quad$\sigma_N, \sigma_B, \sigma_T$ &  Current Needs, Budget, and Time Pressure \\
$\mathcal{H}$ & Dialogue History $\{(r_1, u_1), \dots\}$ \\
\quad$r$ & Response of the Sales Agent \\
\quad$u$ & Response of the User Agent \\
\midrule
\multicolumn{2}{l}{\textit{Cognitive Overload Modeling}} \\
\addlinespace[0.5ex]
$\mathbf{v}$ & Overload vector $[v_a, v_s, v_t, v_u]^\top$ \\
$v_a$ & \textbf{Assortment Size} \\
$v_s$ & \textbf{Choice Set Complexity} \\
\quad $v_{s,d}$ &  Dominance (Presence of dominant option) \\
\quad $v_{s,a}$ &  Alignability (Attribute commensurability) \\
$v_t$ & \textbf{Task Difficulty} \\
\quad $v_{t,p}$ &  Time Pressure \\
\quad $v_{t,a}$ &  Number of Attributes \\
\quad $v_{t,f}$ &  Format Complexity \\
$v_u$ & \textbf{Preference Uncertainty} \\
$\boldsymbol{\beta}$ & Coefficients of the four overload factors \\
$d_{\text{total}}$ & Calculated overload effect size\\
\midrule
\multicolumn{2}{l}{\textit{Decision Process}} \\
\addlinespace[0.5ex]
$\mathcal{I}, \mathcal{C}$ & Initial and candidate recommendation sets ($\mathcal{C} \subseteq \mathcal{I}$) \\
$\theta$ & Cognitive threshold for filtering \\
$\mathbf{a}_i, \mathbf{w}$ & Attribute vector of item $i$ and user weights \\
$u_i$ & Evaluated utility of item $i$ \\
$P_{\text{base}}$ & Base acceptance probability (default: 0.5) \\
$P_{\text{accept}}$ & Final acceptance probability \\
\bottomrule
\end{tabularx}
\caption{Summary of the main notations.}
\label{tab:notations}
\end{table}

\section{Theoretical Background}
\subsection{Rationale for Omitting Partial Overload Factors}
\label{subsec:factors_omitted_rationale}
Although \citet{chernev2015choice} identifies a broader set of determinants, we selectively exclude dimensions that are incompatible with our experimental design. For example, the Decision Goal factor, including decision intent (buying vs.\ browsing) and decision accountability, is omitted. In our setting, user profiles are initialized with explicit goal-oriented purchase needs, rendering browsing behavior irrelevant. Moreover, accountability effects, which arise when consumers must justify their decisions to others, do not apply in a private interaction between a simulated user agent and a sales system. Within our experimental framework, these dimensions inherently yield a null effect; specifically and contribute an effect size of zero to the regression model established by \citet{chernev2015choice}. Consequently, their exclusion does not compromise the validity of the analysis.
\subsection{The Formula Derivation for Equation~\ref{eq:p_accept}}
\label{subsec:d_value_derivation}
Starting from the definition of the total effect size:
\begin{equation}
d_{\text{total}} =
2\left(
\arcsin \sqrt{P_{\text{base}}}
-
\arcsin \sqrt{P_{\text{adjust}}}
\right)
\end{equation}

Dividing both sides by $2$ gives
\begin{equation}
\frac{d_{\text{total}}}{2}
=
\arcsin \sqrt{P_{\text{base}}}
-
\arcsin \sqrt{P_{\text{adjust}}}.
\end{equation}

Rearranging to isolate the term containing $P_{\text{adjust}}$:
\begin{equation}
\arcsin \sqrt{P_{\text{adjust}}}
=
\arcsin \sqrt{P_{\text{base}}}
-
\frac{d_{\text{total}}}{2}.
\end{equation}

Applying the sine function to both sides yields
\begin{equation}
\sqrt{P_{\text{adjust}}}
=
\sin\left(
\arcsin \sqrt{P_{\text{base}}}
-
\frac{d_{\text{total}}}{2}
\right).
\end{equation}

Finally, squaring both sides and letting
$P_{\text{adjust}} = P_{\text{accept}}$ gives
\begin{equation}
P_{\text{accept}}
=
\sin^2\left(
\arcsin \sqrt{P_{\text{base}}}
-
\frac{d_{\text{total}}}{2}
\right).
\end{equation}

\subsection{Design for \texorpdfstring{$P_{\text{base}}$}{Pbase}}

\label{subsec:p_base_design}
Notably, the impact of assortment size on consumer behavior is bi-directional and contingent upon the specific moderating environment. When the antecedents of overload—such as task difficulty or choice set complexity—are at low levels, the observed effect size (Cohen's $d$) typically becomes negative. This negative $d$-value signifies a "more-is-better" effect, where larger assortments provide a competitive advantage by increasing the likelihood of an ideal preference match without exceeding cognitive limits. To achieve behavioral high-fidelity, a simulation framework must therefore support bi-directional dynamics: it must decrease the acceptance probability to model choice deferral under high-overload conditions, while conversely increasing it when the cognitive load is sufficiently low to reflect the utility gains and increased satisfaction inherent in extensive assortments.

\subsection{Connection to the Meta-Analytic Regression Model}
\label{subsec:regression_model}
\citet{chernev2015choice} estimates the magnitude of the choice overload
effect using a meta-analytic regression model of the form

\begin{equation}
d = \beta_0 + \beta_s x_s + \beta_t x_t + \beta_u x_u + \epsilon
\end{equation}

where $d$ denotes the standardized effect size reported in each experimental observation, $x_k$ represents moderator variables characterizing the decision environment, and $\beta$ denotes the regression coefficients estimated in the meta-analysis (see Table~\ref{tab:overload_parameters} for the exact values).
\begin{table}[h]
\centering
\begin{tabular}{lccc}
\toprule
\textbf{Factor} & $\beta_k$ & $\delta_k^{\min}$ & $\delta_k^{\max}$ \\
\midrule
$v_a$ & 0.41 & -0.18 & 1.22 \\
$v_s$ & 0.55 & -1.65 & 0.48 \\
$v_t$ & 0.37 & -0.59 & 0.81 \\
$v_u$ & 0.32 & -1.34 & 1.21 \\
\bottomrule
\end{tabular}
\caption{Parameters used to compute the overload effect size.}
\label{tab:overload_parameters}
\end{table}

In the original analysis, assortment size is primarily operationalized
through experimental contrasts between small and large choice sets and
is therefore absorbed into the intercept term. In our simulation
environment, however, the number of recommended items varies
explicitly across dialogue turns. To capture this variation, we
reparameterize the baseline effect as a moderator associated with
assortment size.

Concretely, we express the assortment contribution as

\begin{equation}
d_a = f_{\text{interp}}(v_a, \delta_a)
\end{equation}

and define the total effect size as

\begin{equation}
d_{total} =
\beta_a d_a +
\beta_s d_s +
\beta_t d_t +
\beta_u d_u
\end{equation}

where the assortment coefficient $\beta_a$ reflects the baseline
magnitude of the assortment-size contrast captured by the intercept
term in the original meta-analytic regression.

\section{Implementation Details}
\label{app:implementation}

\subsection{User Profile Construction}
\label{app:profile_construction}

The global user state $\mathcal{G}$ is defined as the combination of a persona
$\mathcal{P} = \{\phi_O, \phi_K, \phi_U\}$ and a shopping scenario
$\mathcal{S} = \{\sigma_N, \sigma_B, \sigma_T\}$.
We instantiate these components using synthesized profiles derived from the Amazon Reviews 2023 corpus.

\paragraph{Persona and Scenario Instantiation}
\label{app:user_profile}
Following \citet{kim2025towards}, the openness value $\phi_O$ is inferred from historical reviews via LLM-based analysis.
The pickiness $\phi_K$ and preference uncertainty $\phi_U$ are treated as controllable variables in $\{1,2,3\}$.
For the shopping scenario, the dynamic budget $\sigma_B$ is sampled from the interquartile range of prices within the corresponding leaf category to ensure realistic purchasing power, while the time pressure $\sigma_T$ serves as a task-specific constraint.

The granularity of the user’s current needs $\sigma_N$ is modulated by preference uncertainty $\phi_U$.
When uncertainty is medium or high ($\phi_U \in \{2,3\}$), the user is modeled with vague preferences: item descriptions are limited to coarse attributes, and the weight vector $\mathbf{w}$ is initialized uniformly.
Conversely, when uncertainty is low ($\phi_U = 1$), the user is modeled as a more expert decision-maker, using a non-uniform weight vector $\mathbf{w}$ extracted from historical review data.

\paragraph{Asymmetric Initialization}
To better reflect realistic sales interactions, we adopt an asymmetric profile design.
The User Agent is initialized with the complete global state $\mathcal{G}$.
In contrast, the Sales Agent is provided only with the observable profile
$\{\phi_O, \phi_K\}$, representing traits that could plausibly be inferred from a user’s review history.
This information asymmetry requires the Sales Agent to proactively infer the user’s latent intent and constraints through dialogue.

\subsection{Sales Agent Implementation}
\label{app:sales_agent}

\paragraph{Recommendation Retrieval Mechanism}
To facilitate efficient and relevant item retrieval, we construct a vector-based retrieval system using \texttt{Qwen3-Embedding-0.6B} as the backbone embedding model. All retrieval operations are performed using cosine similarity over an HNSW-based vector index, returning the top-$K$ results ($K=50$).

\paragraph{Database Establishment}
For each item in Amazon Reviews 2023 across the target domains, we construct a textual representation by concatenating the title, category path, and product description.
This representation is then embedded into a 1{,}024-dimensional vector space and stored in a vector database for subsequent retrieval.

\paragraph{Retrieval Mechanisms}
The Sales Agent employs two retrieval modes to support different strategic actions.

\begin{itemize}
    \item \textbf{Query-based Retrieval (\textit{Suggest}).}
    When the agent identifies explicit user preferences, it converts the inferred preference description into a query embedding and retrieves the top-$K$ candidate items from valid subcategories.

    \item \textbf{Item-based Retrieval (\textit{Persuade}).}
    To support persuasive strategies, the agent performs a similarity search centered on the currently recommended item. By retrieving items with high semantic similarity, the agent can present alternative products that reinforce salient attributes or provide comparative context.
\end{itemize}

\section{Experimental Setup}
\label{app:exp_details}

\subsection{Evaluation Metrics}
\label{app:metric_details}

\subsubsection{Success Rate}
\label{app:sr_metric}

Success Rate (SR) measures the proportion of simulated sessions that end with a completed purchase decision.
Given $N$ simulated sessions, it is defined as

\begin{equation}
SR = \frac{1}{N} \sum_{i=1}^{N} \mathbf{I}(\text{purchase}_i),
\end{equation}

where $\mathbf{I}(\cdot)$ is the indicator function, which equals $1$ if session $i$ ends in purchase and $0$ otherwise.

\subsubsection{Subjective Dialogue Quality}
\label{app:subjective_metric}

We evaluate dialogue quality using pairwise comparisons between dialogues generated by different user agents.
Given four agents, we construct all $\binom{4}{2}$ agent pairs and conduct dialogue-level comparisons for each pair. Each pair is evaluated along five dimensions: \textit{realism}, \textit{naturalness}, \textit{relevance}, \textit{clarity}, and \textit{adaptability}, where the judge selects the better dialogue for each dimension.

We use \textit{Gemini 3 Flash} as the judge model because it provides stronger analytical and reasoning capabilities than the backbone model, \textit{gpt-oss-20b}, used for dialogue simulation \cite{white2024livebench}.
For each comparison setting, we aggregate the outcomes across all $N=40$ simulated sessions and report the resulting win rates.

\subsubsection{Hallucination Ratio}
\label{app:hr_metric}

To assess the believability of the simulated user agent, we compute the Hallucination Ratio ($\bar{H}_R$), which quantifies the proportion of mentioned entities that are not supported by the available evidence.

For each dialogue turn $t$ in scenario $i$, the turn-level hallucination ratio is defined as

\begin{equation}
H_{R,i,t} =
\frac{H_{i,t}}{H_{i,t} + R_{i,t}},
\end{equation}

where $H_{i,t}$ is the number of hallucinated entities and $R_{i,t}$ is the number of grounded entities.
An entity is considered grounded if it can be verified from either the user profile $\mathcal{P}$ or the dialogue history $\mathcal{H}$; otherwise, it is counted as hallucinated.
Entity extraction and classification are performed using a structured LLM-as-a-judge parser.

To obtain the final reported score, we first average the turn-level ratios over the $T=20$ turns within each scenario, and then average the resulting scenario-level values across all $N=40$ independent sessions:

\begin{equation}
\bar{H}_R =
\frac{1}{N} \sum_{i=1}^{N}
\left(
\frac{1}{T} \sum_{t=1}^{T} H_{R,i,t}
\right).
\end{equation}
user’s historical review sentiments.

\subsection{Hyper-parameters for Overload Experiments}
\label{app:overload_config}

\begin{table}[h]
\centering
\small
\begin{tabular}{lccc}
\toprule
 & \textbf{Low} & \textbf{Medium} & \textbf{Severe} \\
\midrule
$\phi_O$ (Openness) & 2 & 2 & 2 \\
$v_{t,p}$ (Time Pressure) & 1 & 2 & 3 \\
$v_{t,f}$ (Format Complexity) & 1 & 1 & 3 \\
$v_u$ (Preference Uncertainty) & 1 & 2 & 3 \\
\midrule
$|\mathcal{I}|$ (Recommendations) & 3 & 3 & 3 \\
$v_{t,a}$ (Number of Attributes) & 8 & 8 & 8 \\
\bottomrule
\end{tabular}
\caption{Hyper-parameter settings used in the overload-effect experiment.}
\label{tab:overload_hyperparams}
\end{table}






\subsection{Hyper-parameters for Behavioral Reproduction}
\label{app:psycho_repro}

The following parameters are fixed across all experiments:

\begin{table}[ht]
\centering
\small
\begin{tabular}{lc}
\toprule
Parameter & Value \\
\midrule
$\phi_O$ (Openness) & 2 \\
$\phi_K$ (Pickiness) & 2 \\
$v_{t,p}$ (Time Pressure) & 2 \\
$v_{t,f}$ (Format Complexity) & 1 \\
\bottomrule
\end{tabular}
\caption{Fixed hyper-parameters in behavioral reproduction experiments.}
\end{table}

\paragraph{Total Information Curve}
Both assortment size $v_a$ and number of attributes $v_{t,a}$ are varied.

\paragraph{Assortment Size Curve}
$v_{t,a}=5$, while $v_a$ is varied.

\paragraph{Number of Attributes Curve}
$v_a=3$, while $v_{t,a}$ is varied.

\section{Use-of-LLMs}
Large language models (LLMs) are used in this work primarily as core components of the proposed framework. Specifically, LLMs serve as the backbone of the user simulator, generating conversational behaviors under different experimental conditions. They are also used as embedding models to encode items into dense representations for the embedding process, and as automatic evaluators (LLM-as-a-judge) to provide complementary evaluation signals.

Outside the modeling pipeline, LLMs are used only for minor writing and coding assistance, without affecting the experimental design or results.
\section{Additional Experimental Results}
\label{app:additional_results}

\subsection{Additional Overload Results Across Sales-Agent Settings}
\label{app:exp_var}

\begin{figure*}[ht]
  \centering
  \includegraphics[width=\textwidth]{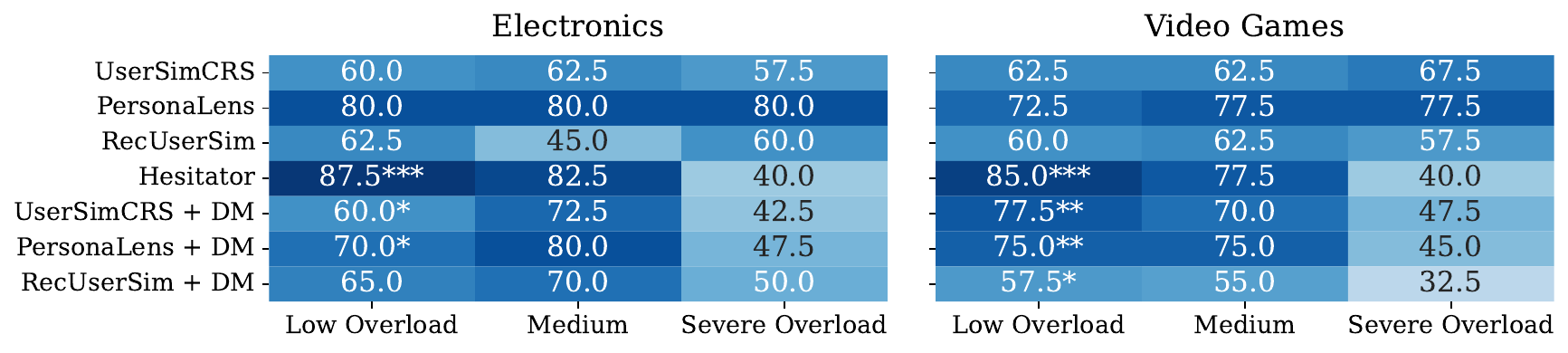}
  \caption{Simulation success rates under varying levels of cognitive overload when the Sales Agent operates in persuasive mode. The overall trend remains consistent with the main findings: after adding the Decision Module, user simulators exhibit a clearer decline in success rates as overload increases.}
  \label{fig:persuasive_overload}
\end{figure*}

\begin{figure*}[ht]
  \centering
  \includegraphics[width=\textwidth]{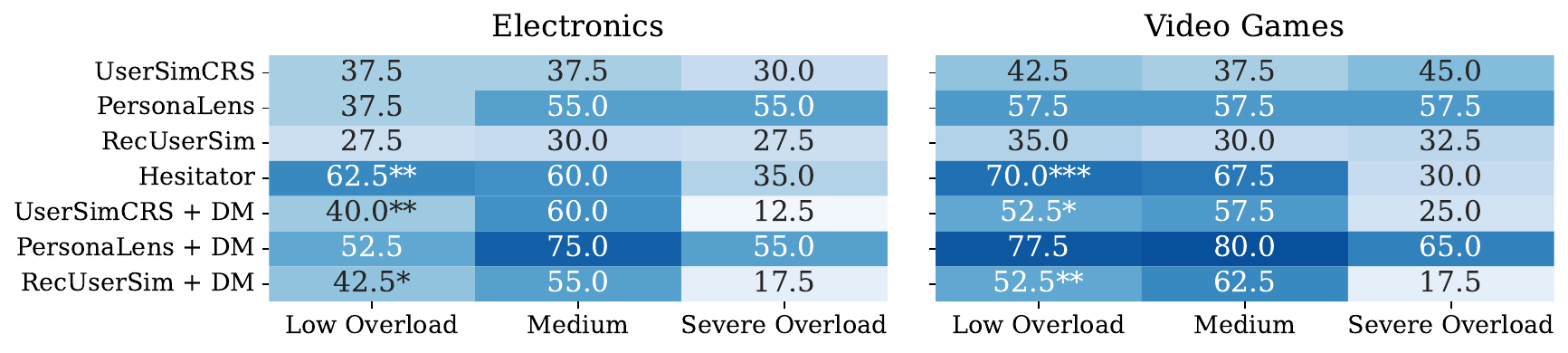}
  \caption{Simulation success rates under varying levels of cognitive overload across different LLM backbones. The Decision Module consistently improves overload sensitivity across backbone choices, indicating that the behavioral correction effect is robust to the underlying model.}
  \label{fig:backbone_overload}
\end{figure*}

\end{document}